# Geographically and temporally weighted neural networks for satellite-based mapping of ground-level PM$_{2.5}$

Tongwen Li [a], Huanfeng Shen [a,b,c*], Qiangqiang Yuan [d,b], Liangpei Zhang [e,b]

[a] School of Resource and Environmental Sciences, Wuhan University, Wuhan, 430079, China.

[b] Collaborative Innovation Center of Geospatial Technology, Wuhan, 430079, China.

[c] The Key Laboratory of Geographic Information System, Ministry of Education, Wuhan University, Wuhan, 430079, China.

[d] School of Geodesy and Geomatics, Wuhan University, Wuhan, 430079, China.

[e] The State Key Laboratory of Information Engineering in Surveying, Mapping and Remote Sensing, Wuhan University, Wuhan, 430079, China.

\* Corresponding author. E-mail address: shenhf@whu.edu.cn

**Highlights**

- GTWNNs were developed for satellite-based estimation of ground-level PM$_{2.5}$.
- Nonlinearity and spatiotemporal heterogeneities were simultaneously considered.
- The proposed GTWNNs model outperformed previous spatiotemporal models.

**Abstract**

The integration of satellite-derived aerosol optical depth (AOD) and station-measured $PM_{2.5}$ provides a promising approach for obtaining spatial $PM_{2.5}$ data. Several spatiotemporal models, which considered spatial and temporal heterogeneities of AOD-$PM_{2.5}$ relationship, have been widely adopted for $PM_{2.5}$ estimation. However, they generally described the complex AOD-$PM_{2.5}$ relationship based on a linear hypothesis. Previous machine learning models yielded great superiorities for fitting the nonlinear AOD-$PM_{2.5}$ relationship, but seldom allowed for its spatiotemporal variations. To simultaneously consider the nonlinearity and spatiotemporal heterogeneities of AOD-$PM_{2.5}$ relationship, geographically and temporally weighted neural networks (GTWNNs) were developed for satellite-based estimation of ground-level $PM_{2.5}$ in this study. Using satellite AOD products, NDVI data, and meteorological factors over China as input, GTWNNs were set up with station $PM_{2.5}$ measurements. Then the spatial $PM_{2.5}$ data of those locations with no ground stations could be obtained. The proposed GTWNNs have achieved a better performance compared with previous spatiotemporal models, i.e., daily geographically weighted regression and geographically and temporally weighted regression. The sample-based and site-based cross-validation $R^2$ values of GTWNNs are 0.80 and 0.79, respectively. On this basis, the spatial $PM_{2.5}$ data with a resolution of 0.1 degree were generated in China. This study implemented the combination of geographical law and neural networks, and improved the accuracy of satellite-based $PM_{2.5}$ estimation.

# 1. Introduction

Fine particulate matter (PM$_{2.5}$), which refers to particulate matters with aerodynamic diameters of less than 2.5 $\mu m$, has been reported to have many adverse impacts on climate change (Cao et al., 2012; Zhao et al., 2006), human health (Guo et al., 2016; Wu et al., 2017), and so on. Consequently, PM$_{2.5}$ pollution has become a hotspot issue of common concern for the public in recent years (Peng et al., 2016; van Donkelaar et al., 2016). However, currently, the monitoring of PM$_{2.5}$ pollution is still limited due to the sparely distributed ground stations. Given this, it is an urgent need to obtain spatial PM$_{2.5}$ data with a high resolution for dynamic monitoring and control of atmospheric PM$_{2.5}$ pollution (Engel-Cox et al., 2013).

Because satellite remote sensing is able to provide observations with large temporal and spatial coverages, it has been extensively employed for the monitoring of ground-level PM$_{2.5}$ (Hoff and Christopher, 2009; Lary et al., 2014; Liu et al., 2005; Liu et al., 2009; Martin, 2008; Zhan et al., 2017). The most widely used parameter is satellite-derived aerosol optical depth (AOD). To estimate ground-level PM$_{2.5}$ from satellite-derived AOD, a typical strategy used is to establish the statistical relationship between them (the AOD-PM$_{2.5}$ relationship). The simplest method is to establish a linear AOD-PM$_{2.5}$ relationship using a linear regression (LR) model (Chu et al., 2003; Wang and Christopher, 2003). However, the AOD-PM$_{2.5}$ relationship is easily influenced by many other factors, such as meteorological conditions. Then, the multiple linear regression (MLR) models were applied to estimate PM$_{2.5}$ from satellite-derived AOD (Gupta and Christopher, 2009). Unlike MLR model, some numeric priors (e.g., exponential relationship) were introduced into the AOD-PM$_{2.5}$ model, the semi-empirical model (SEM) was developed for satellite-based PM$_{2.5}$ estimation (Liu et al., 2005; Tian and

Chen, 2010). These early models usually established a constant AOD-PM$_{2.5}$ relationship in the whole study region and period.

However, the AOD-PM$_{2.5}$ relationship tends to vary with time and locations, so some spatially and/or temporally varying models have been developed for AOD-based estimation of PM$_{2.5}$. To consider the temporal variations of AOD-PM$_{2.5}$ relationship, linear mixed effect (LME) models often add a random effect on each time (e.g., day, month) (Lee et al., 2011; Ma et al., 2016a). Then, the ground-level PM$_{2.5}$ are estimated using time-specific coefficients. Additionally, geographically weighted regression (GWR) models, which take the spatial heterogeneity of AOD-PM$_{2.5}$ relationship into consideration, have been widely applied for the estimation of PM$_{2.5}$ (Hu et al., 2013; Ma et al., 2014). Through a local regression technique, the GWR model estimates PM$_{2.5}$ concentrations from satellite AOD using spatially varying coefficients. Then, the temporal dependency of AOD-PM$_{2.5}$ relationship was further considered, geographically and temporally weighted regression (GTWR) models were introduced for the AOD-PM$_{2.5}$ modeling (Bai et al., 2016; Guo et al., 2017; He and Huang, 2018a). Due to the simultaneous use of spatial and temporal information, the GTWR model is expected to outperform the GWR model. However, these widely used spatiotemporally varying models describe the AOD-PM$_{2.5}$ relationship based on a linear hypothesis, while the AOD-PM$_{2.5}$ relationship has been proven to be very complicated (e.g., nonlinear) (Yang et al., 2018; Zheng et al., 2017).

To fit nonlinear relationships between PM$_{2.5}$ and influencing factors, the machine learning models have shown notable advantages than traditional statistical models (Li et al., 2017b). Consequently, many machine learning methods have been applied for the AOD-PM$_{2.5}$ modeling, such as neural networks (Di et al., 2016; Li et al., 2017b), random forest (Hu et al., 2017; Zhan et al., 2018), gradient boosting machine (Zhan et al., 2017), and so on. However, previous studies using machine learning models often established a globally numeric AOD-PM$_{2.5}$ relationship, and neglected the spatial and temporal heterogeneities of AOD-PM$_{2.5}$ relationship. The ground-level PM$_{2.5}$ were estimated from satellite AOD using a constant model for the whole study region and period. As a result, these models tended to obtain a good result as a whole, but might fail to get satisfactory estimates for a specific location or time.

It can be found from above analyses that two main modeling strategies, i.e., global modeling and spatiotemporal modeling, have been adopted to describe the AOD-PM$_{2.5}$ relationship. The global modeling (e.g., LR, MLR, SEM, machine learning models) does not vary with time and space, and represents the AOD-PM$_{2.5}$ relationship using constant coefficients in the whole study region and period. Unlike global modeling, the spatiotemporal modeling (e.g., LME, GWR, GTWR) establishes the AOD-PM$_{2.5}$ relationship using time-specific and/or location-specific coefficients, which actually considers the spatial and temporal variations of AOD-PM$_{2.5}$ relationship. Hence, the spatiotemporal modeling has become popular approaches for the AOD-PM$_{2.5}$ modeling. However, they generally describe the complex AOD-PM$_{2.5}$ relationship based on a linear hypothesis. To fit the nonlinear relationships between PM$_{2.5}$ and influencing factors, machine learning models have shown notable advantages. While the global machine learning modeling seldom considers the spatial and temporal variations of the AOD-PM$_{2.5}$ relationship. Therefore, a challenging proposition is whether we can incorporate spatial and temporal heterogeneities into machine learning models for establishing spatiotemporally varying nonlinear AOD-PM$_{2.5}$ relationships.

The main objective of this work is to develop geographically and temporally weighted neural networks (GTWNNs), which simultaneously consider the nonlinearity and spatiotemporal heterogeneities of AOD-PM$_{2.5}$ relationship, for the satellite-based estimation of ground-level PM$_{2.5}$ in China. With the input of satellite observations and meteorological factors, and the output of ground-level PM$_{2.5}$ concentrations, the GTWNNs will be established for the modeling of AOD-PM$_{2.5}$ relationship. To fully verify the feasibility of GTWNNs, the daily GWR (D-GWR) and GTWR models will be included for comparisons. This study will investigate and validate the effectiveness of combining geographical law and machine learning, and provide new approaches for the satellite-based estimation of ground-level PM$_{2.5}$.

## 2. Study region and data

### 2.1. Ground-level PM$_{2.5}$ measurements

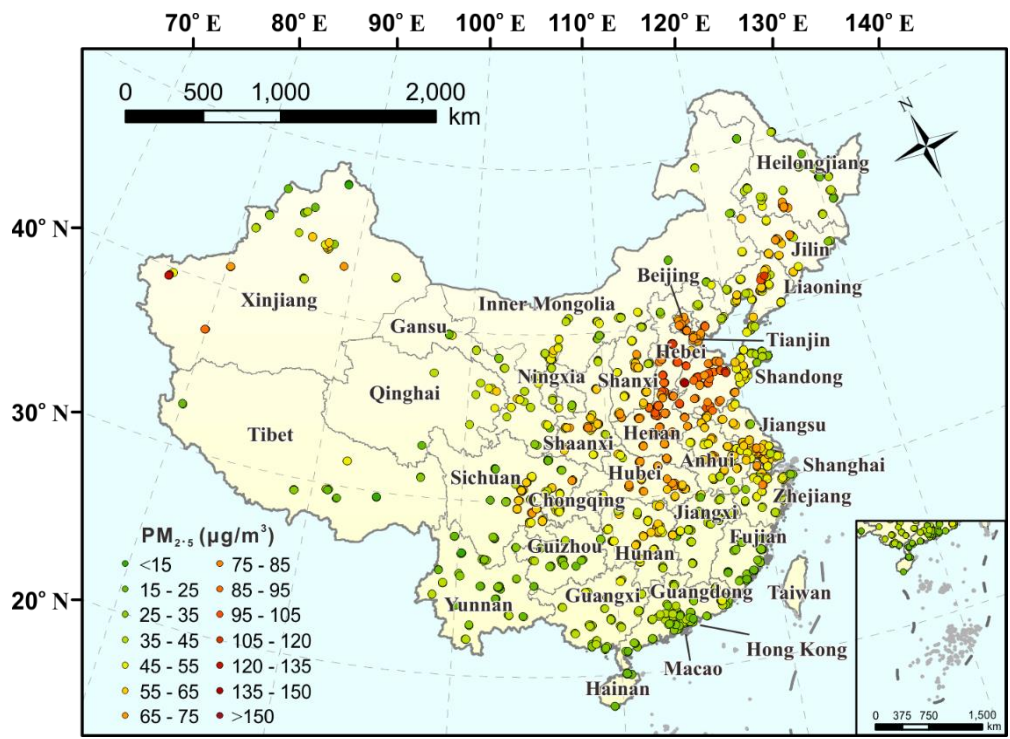

**Fig. 1**. Spatial distribution of ground PM$_{2.5}$ stations in China.

The study region of this paper is China, which is shown in Fig. 1. The hourly PM$_{2.5}$

measurements from 1 January 2015 to 31 December 2015 were collected from the China National Environmental Monitoring Center (CNEMC) Web site (http://106.37.208.233:20035/). By the end of 2015, a total of ~1500 stations have been established for the monitoring of atmospheric pollution. The hourly $PM_{2.5}$ measurements were averaged to daily mean $PM_{2.5}$, with the motivations reported in our previous study (Shen et al., 2017). In this work, only those days containing more than 18 valid hourly measurements were included for the model development. The annual mean values of $PM_{2.5}$ concentration for each station were calculated and presented in Fig. 1.

*2.2. Satellite observations*

The AOD products obtained from Moderate Resolution Imaging Spectroradiometer (MODIS) onboard Terra and Aqua satellite were widely used for the estimation of ground-level $PM_{2.5}$ (Wang et al., 2018). Both Terra AOD and Aqua AOD products were included in this paper, which were downloaded from the Level 1 and Atmosphere Archive and Distribution System (LAADS, https://ladsweb.modaps.eosdis.nasa.gov/). We adopted AOD products of collection 6 with a spatial resolution of 10 km, and extracted the data field of "AOD_550_Dark_Target_Deep_Blue_Combined" to establish the AOD-$PM_{2.5}$ model. This AOD products are retrieved through the combination of dark target and deep blue algorithms (Levy et al., 2013). To estimate daily $PM_{2.5}$ concentration, the average of Terra AOD and Aqua AOD were exploited. For each pixel, if only Terra AOD or Aqua AOD is available, then it is selected; if both of them are available, their average is used for the estimation of $PM_{2.5}$. However, if both of them are absent, this pixel is identified to be missing and cannot provide $PM_{2.5}$ retrievals.

Additionally, MODIS normalized difference vegetation index (NDVI, MOD13) products were also downloaded from the LAADS website. The MODIS NDVI data are available at a spatial resolution of 1 km every 16 days. NDVI data was used in the AOD-PM$_{2.5}$ model to reflect the land-use type.

*2.3. Meteorological variables*

Since the meteorological factors were found to have influences on PM$_{2.5}$ concentrations (Yang et al., 2017), we obtained meteorological variables from the National Aeronautics and Space Administration (NASA) atmospheric reanalysis data known as the second Modern-Era Retrospective Analysis for Research and Applications (MERRA-2) (Molod et al., 2015). They have a spatial resolution of 0.625° longitude × 0.5° latitude. The meteorological variables used in this study include as follows: wind speed at 10 m above ground (WS, m/s), air temperature at a 2 m height (TMP, K), relative humidity (RH, %), surface pressure (PS, kPa), and planetary boundary layer height (PBL, m). To match with the satellite observations, the meteorological variables corresponding to the satellite overpass time (about 9:00~16:00 China standard time) were averaged for the model establishment. We refer the readers to the official website (http://gmao.gsfc.nasa.gov/GMAO_products/) for more details.

*2.4. Data preprocessing and matching*

Before model development, the data preprocessing were conducted to obtain a temporally and spatially consistent dataset. Firstly, all the data were reprojected to the same projection coordinate system. Secondly, satellite AOD products, meteorological data, and satellite NDVI were resampled to 0.1 degree using the bilinear method. Finally, we extracted the satellite data and meteorological data on those locations where the PM$_{2.5}$ ground stations locate. For each

0.1-degree grid cell, ground PM$_{2.5}$ measurements from multiple stations were averaged. After the data preprocessing and matching, in total, 64458 records were obtained as the fundamental dataset for the model establishment and validation.

**3. Geographically and temporally weighted neural networks for the estimation of PM$_{2.5}$**

*3.1. Model development*

In this study, the spatial and temporal variations of AOD-PM$_{2.5}$ relationship are simultaneously taken into consideration in the geographically and temporally weighted neural networks. The structure of this model is depicted as Eq. (1).

$$PM_{2.5j} = f_{(x_j, y_j, t_j)}\left(AOD_j, RH_j, WS_j, TMP_j, PS_j, PBL_j, NDVI_j\right) \qquad (1)$$

Where $PM_{2.5j}$ refers to the daily mean PM$_{2.5}$ concentration of prediction grid cell $j$, which is at the location of $(x_j, y_j)$ and on the day of $t_j$. Here, $(x_j, y_j)$ is the central coordinates of a grid cell, and $t_j$ is the day of year. $f_{(x_j, y_j, t_j)}(\ )$ denotes the location-time-specific estimation function for ground-level PM$_{2.5}$ using input variables. That is, the estimation function varies from location to location and day to day. Specifically, this estimation function is represented using a generalized regression neural network (GRNN) model (Specht, 1991; Specht, 1993), and Fig. 2 shows the schematic of geographically and temporally weighted GRNN (GTWGRNN) model for the estimation of PM$_{2.5}$.

As presented in Fig. 2, a common GRNN architecture consists of three layers. The first layer is input layer, in which the variables of satellite AOD, RH, WS, TMP, PS, PBL and NDVI are incorporated. The second layer is hidden layer, which has as many neurons as the number of samples used for constructing the GRNN model. The last layer has only one node, i.e., ground-level PM$_{2.5}$ concentration. Here, the GRNN model is exploited to create a continuous surface of PM$_{2.5}$ with the input variables. The GRNN model has been proven to inherently have the advantages of fast converging to a global minimum (Cigizoglu and Alp, 2006), therefore it is often applied in the function approximation problems. In addition, unlike the traditional neural networks, less parameters are needed for the GRNN model. Only one parameter namely spread, which controls the smoothness of fitting function, should be adjusted according to the model performance. More details of GRNN model can refer to previous studies (Specht, 1991; Specht, 1993).

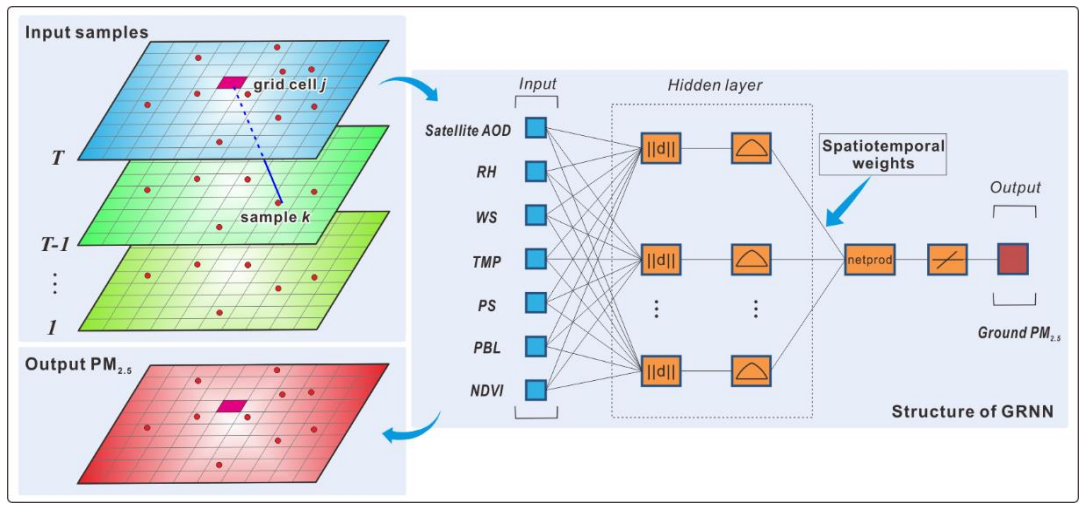

**Fig. 2**. Schematic of the GTWGRNN model for the estimation of PM$_{2.5}$.

In consideration of spatial and temporal variations of AOD-PM$_{2.5}$ relationship, the geographical and temporal weighting is introduced in the GRNN model, that is, the GTWGRNN model. To establish the GTWGRNN model for the prediction grid cell $j$ with the location of $(x_j, y_j)$ and the time of $t_j$ ($t_j = T$, see Fig. 2), the samples on day $T$ and previous days were adopted. Additionally, a popular Gaussian distance decay-based weighting function is introduced to account for the importance of sample $k$ to prediction grid cell $j$ based on the spatial and temporal dimensions (He and Huang, 2018b). The weight of sample $k$ is shown as Eq. (2).

$$w_{jk} = e^{-\frac{(ds_{jk})^2 + \lambda(dt_{jk})^2}{h_{ST}}} \qquad (2)$$

Where $ds$ and $dt$ denote spatial and temporal distance between sample $k$ and prediction grid cell $j$, respectively. There are two essential parameters in the weighting function, one is $\lambda$, which is used as a scale factor to balance the effects of spatial and temporal distances; The other is $h_{ST}$, and it is considered as bandwidth to make the influence of sample $k$ on prediction grid cell $j$ falling into the decay based on the spatiotemporal distance. When $\lambda = 0$, the temporal distance has no influence on the weights, indicating that even the samples on the days far away from day $T$ are weighed based on the spatial distance. When $\lambda = \infty$, only the samples on day $T$ have influences on the model for prediction grid cell $j$, and it is actually a GWGRNN model in this situation. In this study, to save the computation cost, only those samples whose weights were greater than 1E-6 were chosen for the establishment of GTWGRNN model.

Specifically, supposing that $N$ samples are collected for the establishment of GTWGRNN model for prediction grid cell $j$. Let $\mathbf{X}_{M \times N}$ be the input matrix, where $M$ is the number of input variables. Thus, the number of neurons in the hidden layer is $N$. The NN-weights (NN-weight is used here for neural network to distinguish with spatiotemporal weight) between input layer and hidden layer is $\mathbf{X}^T$. Based on the weighting function (Eq. (2)), the samples, which are more spatially and temporally closer to prediction grid cell $j$, show larger contributions to the output PM$_{2.5}$. Incorporating spatiotemporal weights into GRNN, the NN-weight between the $ith$ neuron in the hidden layer and the output neuron can be solved. Thus, the PM$_{2.5}$ concentration on the prediction cell $j$ can be inferred with the input variables and the NN-weights of neural network (NN-weights between input layer and hidden layer, and between hidden layer and output layer).

*3.2. Parameter selection of GTWGRNN model*

There are totally three essential parameters in the GTWGRNN model, i.e., the scale factor ($\lambda$), the spatiotemporal bandwidth ($h_{ST}$), and the spread of GRNN. In line with previous study (He and Huang, 2018b), a cross-validation technique was adopted to select the parameters of the scale factor ($\lambda$) and the spatiotemporal bandwidth ($h_{ST}$). Through a nested-loop process, the values of $\lambda$ and $h_{ST}$ were choses as 6E3 and $120\, km \cdot day$ respectively. Additionally, two weighting regimes (fixed bandwidth and adaptive bandwidth) are optional. For the fixed bandwidth regime, distance is constant but the number of nearest neighbors varies. For the adaptive bandwidth regime, the number of nearest neighbors remains constant but the distance varies. In this study, we adopted a Gaussian distance decay-based weighting function for the adaptive bandwidth regime, indicating that samples outside the neighborhood were not cut off. With a view to model accuracy and computation complexity, the adaptive bandwidth regime ($\lambda$ =3E4 and $h_{ST}$ =4 (the average of spatiotemporal bandwidth is $144\, km \cdot day$)) were finally selected for the whole China. The reason for this is that it can achieve a competitive performance compared with the regime of fixed bandwidth, and obtain robust mapping results of $PM_{2.5}$ concentrations more effectively with the uneven distribution of $PM_{2.5}$ stations. For the spread of GRNN, it is a key parameter to influence the smoothness of fitting function. We tested the parameter of spread with bounds of 0.01 to 0.5, and the results suggested that the model with spread of 0.1 performed the best for the estimation of $PM_{2.5}$ concentrations.

*3.3. Model evaluation*

To evaluate the model predictive ability, a 10 fold cross-validation technique (Rodriguez et al., 2010) has been extensively adopted. For the satellite-based mapping of $PM_{2.5}$, there are mainly two cross-validation strategies, i.e., sample-based cross-validation and site-based cross-validation. For the sample-based cross-validation, all the collected samples were divided into 10 folds randomly with approximately equal number of records in each fold. Secondly, nine folds of them were exploited for model fitting, and the remaining one fold was used for validation. Finally, the above step would be repeated ten times so as to validate the model performance on each fold of validation samples. For the site-based cross-validation, the grid cells containing monitoring sites were divided into 10 folds, and then, the validation process was similar with sample-based cross-validation. However, the input data in validation set was monitoring sites rather than samples. In short, for site-based cross-validation, the validation stations are not used for model fitting all along; for the sample-based cross-validation, the samples on some days may be used for model fitting and the samples on the other days used for validation. That is the main difference between these two validation strategies. Therefore, the sample-based cross-validation reflect the overall predictive ability of the AOD-$PM_{2.5}$ models, and the site-based cross-validation (which is a spatial hold-out strategy), can assess the spatial prediction performance more reasonably. Consistent to our previous studies (Li et al., 2017a; Shen et al., 2017), the sample-based and site-based cross-validation strategies were both employed in this paper. In fact, the "site" here refers to "grid cell", because the ground $PM_{2.5}$ measurements from multiple stations were averaged in a grid cell.

The statistical indicators used included the following four terms: determination coefficient, i.e., R square ($R^2$, unitless), root-mean-square error (RMSE, $\mu g/m^3$), mean predictive error (MPE, $\mu g/m^3$), and relative prediction error (RPE, defined as RMSE divided by the average of ground-level $PM_{2.5}$ with the unit of 100%).

## 4. Results and discussion

*4.1. Descriptive statistics*

Table 1 presents the descriptive statistics of samples in the dataset for the GTWGRNN modeling. The average of ground-level $PM_{2.5}$ concentration is 55 $\mu g/m^3$, with the bounds of 1 to 574 $\mu g/m^3$ and a standard deviation (Std. Dev) of 39 $\mu g/m^3$. The AOD data has minimum and maximum values of -0.05 and 4.76, respectively. The mean value of AOD in the sample dataset is 0.52. The correlation coefficient between AOD and $PM_{2.5}$ is 0.47, which reports a similar level to that in previous studies at a national scale in China (Fang et al., 2016; He and Huang, 2018b).

Table 1 Descriptive statistics of samples in the dataset for the GTWGRNN modeling.

|  | $PM_{2.5}$ ($\mu g/m^3$) | AOD (unitless) | RH (%) | WS (m/s) | TMP (K) | PS (kPa) | PBL (m) | NDVI (unitless) |
|---|---|---|---|---|---|---|---|---|
| Min | 1 | -0.05 | 4.10 | 0.01 | 255.31 | 54.88 | 63.80 | -0.30 |
| Max | 574 | 4.76 | 95.50 | 16.04 | 315.07 | 103.85 | 3927.20 | 0.92 |
| Mean | 55 | 0.52 | 38.99 | 3.85 | 292.01 | 95.24 | 1552.70 | 0.35 |
| Std. Dev | 39 | 0.45 | 14.50 | 2.29 | 10.31 | 7.46 | 572.12 | 0.19 |

*4.2. Evaluation of GTWGRNN model performance*

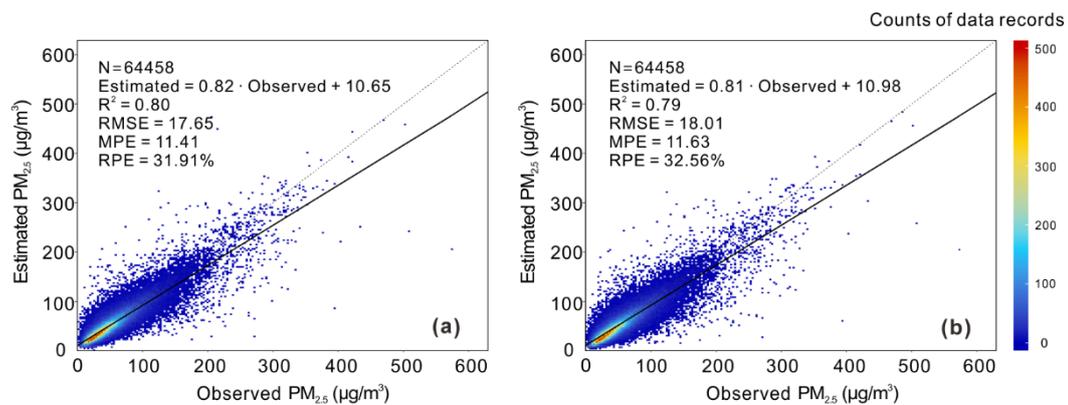

**Fig. 3**. Scatter plots for (a) sample-based and (b) site-based cross-validation results.

As presented in Fig. 3, the proposed GTWGRNN model achieves a satisfactory performance in the cross-validation. For the sample-based cross-validation, the respective $R^2$ and RMSE values are 0.80 and 17.65 $\mu g/m^3$. The results suggest that the GTWGRNN model is able to account for 80% of daily PM$_{2.5}$ variations, and shows a great overall predictive ability for ground-level PM$_{2.5}$ concentrations. Compared with sample-based cross-validation, the site-based cross-validation displays a slight decreasing trend in model performance. The $R^2$ and RMSE values are 0.79 and 18.01 $\mu g/m^3$, respectively. As a spatial hold-out validation strategy, the site-based cross-validation implies that the GTWGRNN model has a robust predictive power for the spatial mapping of PM$_{2.5}$ concentration. Additionally, the slopes of sample-based cross-validation and site-based cross-validation are 0.82 and 0.81, respectively. Although it means the GTWGRNN model tends to underestimate high values and overestimate low values, however this is a common problem in the field of satellite-based PM$_{2.5}$ estimation. Compared with previous studies on AOD-based PM$_{2.5}$ estimation at a national scale in China (Fang et al., 2016), our study still shows slighter extent of underestimation/overestimation for the prediction of ground-level PM$_{2.5}$.

Table 2 shows seasonal performance of the GTWGRNN model. Among four seasons, winter (December, January, and February) obtains the highest $R^2$ values of 0.81 and 0.80 for sample-based and site-based cross-validation, respectively. However, the RMSE values for winter are also the highest, the reason for this is that winter has higher levels of PM$_{2.5}$ concentrations than the other three seasons. Then, both spring (March, April, and May) and autumn (September, October, and November) have achieved $R^2$ values of 0.75 and 0.74 for sample-based and site-based cross-validation, respectively. However, summer (June, July, and August) performs the worst with sample-based cross-validation $R^2$ of 0.68, which agrees with previous findings (He and Huang, 2018b).

Table 2 Seasonal performance of the GTWGRNN model.

| Season | Sample-based cross-validation | | | | Site-based cross-validation | | | |
|---|---|---|---|---|---|---|---|---|
| | $R^2$ | RMSE | MPE | RPE | $R^2$ | RMSE | MPE | RPE |
| Spring | 0.75 | 15.44 | 10.10 | 29.64 | 0.74 | 15.64 | 10.28 | 30.02 |
| Summer | 0.68 | 13.91 | 9.34 | 34.07 | 0.66 | 14.36 | 9.51 | 35.17 |
| Autumn | 0.75 | 17.82 | 11.46 | 35.04 | 0.74 | 18.26 | 11.75 | 35.91 |
| Winter | 0.81 | 22.61 | 15.02 | 29.08 | 0.80 | 23.01 | 15.23 | 29.59 |

Furthermore, the spatial performance of the GTWGRNN model was evaluated, which are presented in Fig. 4. Overall, the GTWGRNN model shows a better performance in the middle and eastern China, whereas a relatively poorer result is reported in the western China. The reason for this could be that the monitoring stations are more densely distributed in the middle and eastern China and sparsely distributed in the western China. Though an uneven distribution of model performance in space, the GTWGRNN model has generally achieves a satisfactory predictive performance, with 71% of grid cells reporting high $R^2$ values of greater than 0.7 for the site-based cross-validation.

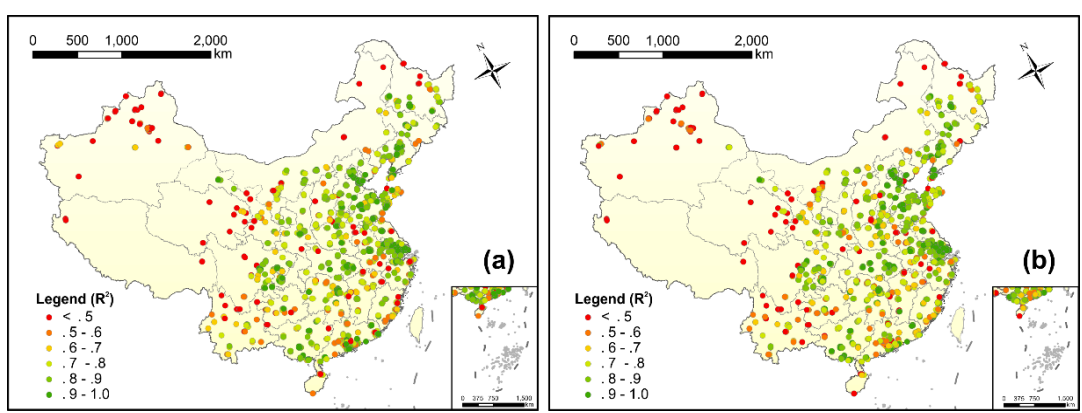

**Fig. 4**. Spatial performance of (a) sample-based cross-validation $R^2$ and (b) site-based cross-validation $R^2$ of the GTWGRNN model.

*4.3. Comparison with other models*

Table 3 presents the sample-based and site-based cross-validation performance of various models. Firstly, to consider the spatial and temporal variations of AOD-PM$_{2.5}$ relationship, the original GRNN (Ori-GRNN) models are established for each location and time using local samples. However, the geographical and temporal weighting is not considered. The Ori-GRNN model performs much worse than GTWGRNN. The sample-based and site-based cross-validation $R^2$ values are 0.72 and 0.71, respectively. The results indicate that incorporating geographical and temporal weighting can boost the accuracy of $PM_{2.5}$ estimation for neural networks. Secondly, compared with the GTWGRNN model, the GWGRNN model only use spatial information for represent the AOD-$PM_{2.5}$ relationship. As a result, a relatively poorer result has been reported, with both sample-based and site-based cross-validation $R^2$ values of 0.78. Through the simultaneous use of spatial and temporal information, the GTWGRNN model achieves a better performance, with sample-based and site-based cross-validation $R^2$ values of 0.80 and 0.79, respectively.

Table 3 Performance of various models.

| Model | Sample-based cross-validation | | | | Site-based cross-validation | | | |
| --- | --- | --- | --- | --- | --- | --- | --- | --- |
| | $R^2$ | RMSE | MPE | RPE | $R^2$ | RMSE | MPE | RPE |
| GTWGRNN | 0.80 | 17.65 | 11.41 | 31.91 | 0.79 | 18.01 | 11.63 | 32.56 |
| GWGRNN | 0.78 | 18.47 | 11.73 | 33.39 | 0.78 | 18.38 | 11.74 | 33.23 |
| Ori-GRNN | 0.72 | 20.39 | 13.64 | 36.79 | 0.71 | 20.58 | 13.80 | 37.13 |
| GTWR | 0.74 | 20.18 | 12.78 | 36.31 | 0.70 | 21.52 | 13.53 | 38.67 |
| D-GWR | 0.69 | 22.17 | 13.52 | 39.68 | 0.69 | 22.12 | 13.51 | 39.54 |

Then follows the comparisons between GTWGRNN, GTWR, and daily GWR (D-GWR) models. Among these three models, the D-GWR model performs relatively worse, with sample-based and site-based cross-validation $R^2$ values of 0.69 and 0.69, respectively. The reason for this could be that the D-GWR model describes the AOD-$PM_{2.5}$ relationship based on a linear hypothesis, and only uses spatial information for the model establishment. By further incorporating the temporal relation of AOD-$PM_{2.5}$ relationship, the GTWR model achieves some advantages than the D-GWR model. However, it should be noted that the site-based cross-validation $R^2$ of GTWR (0.70) reports only a slight superiority than that of D-GWR model (0.69). This finding implies that the samples from adjacent grid cells on previous days actually do not contribute greatly to the estimation of $PM_{2.5}$ at a daily scale. Based on the hypothesis of nonlinearity, the GTWGRNN model has obtained the best performance for both sample-based and site-based cross-validation. These findings reveal the superiority of the GTWGRNN model for the satellite-based estimation of ground-level $PM_{2.5}$.

*4.4. Mapping of ground-level $PM_{2.5}$ concentration*

Based on the proposed GTWGRNN model, the spatial concentrations of $PM_{2.5}$ were predicted. Fig. 5 shows the annual and seasonal distributions of ground-level $PM_{2.5}$, which were mapped using the merging strategy in our previous study (Li et al., 2017b). Overall, the mean value of $PM_{2.5}$ concentration over the whole China is 43.43 $\mu g/m^3$, which exceeds 24% of the China's National Air Quality Standard (CNNAQS) level 2 of 35 $\mu g/m^3$. In addition, it is predicted about 70% of grids cells in China have annual $PM_{2.5}$ concentrations of greater than 35 $\mu g/m^3$. These findings indicate that China is still suffering serious $PM_{2.5}$ pollution, and also show that the integration of satellite remote sensing and station measurements has the potential to provide more spatiotemporal information beyond the ground station networks.

Spatially, the distribution patterns of $PM_{2.5}$ concentration agree with the degree of economic development and urbanization in China. That is, relatively low $PM_{2.5}$ values are found in the western China, whereas the eastern China has higher levels of $PM_{2.5}$ concentrations. One of the most polluted areas is located in the North China Plain, where the annual $PM_{2.5}$ values are generally higher than 75 $\mu g/m^3$. Another pollution center of high $PM_{2.5}$ values is southern

Xinjiang Autonomous region. A possible reason for this is that the accumulation of $PM_{2.5}$ results from the dust particles in this desert region. Seasonally, summer is the cleanest, whereas winter is the most polluted season. According to previous works (Zhang and Cao, 2015), the heavy pollution in winter is possibly caused by winter heating, especially in the regions of north China and northwest China.

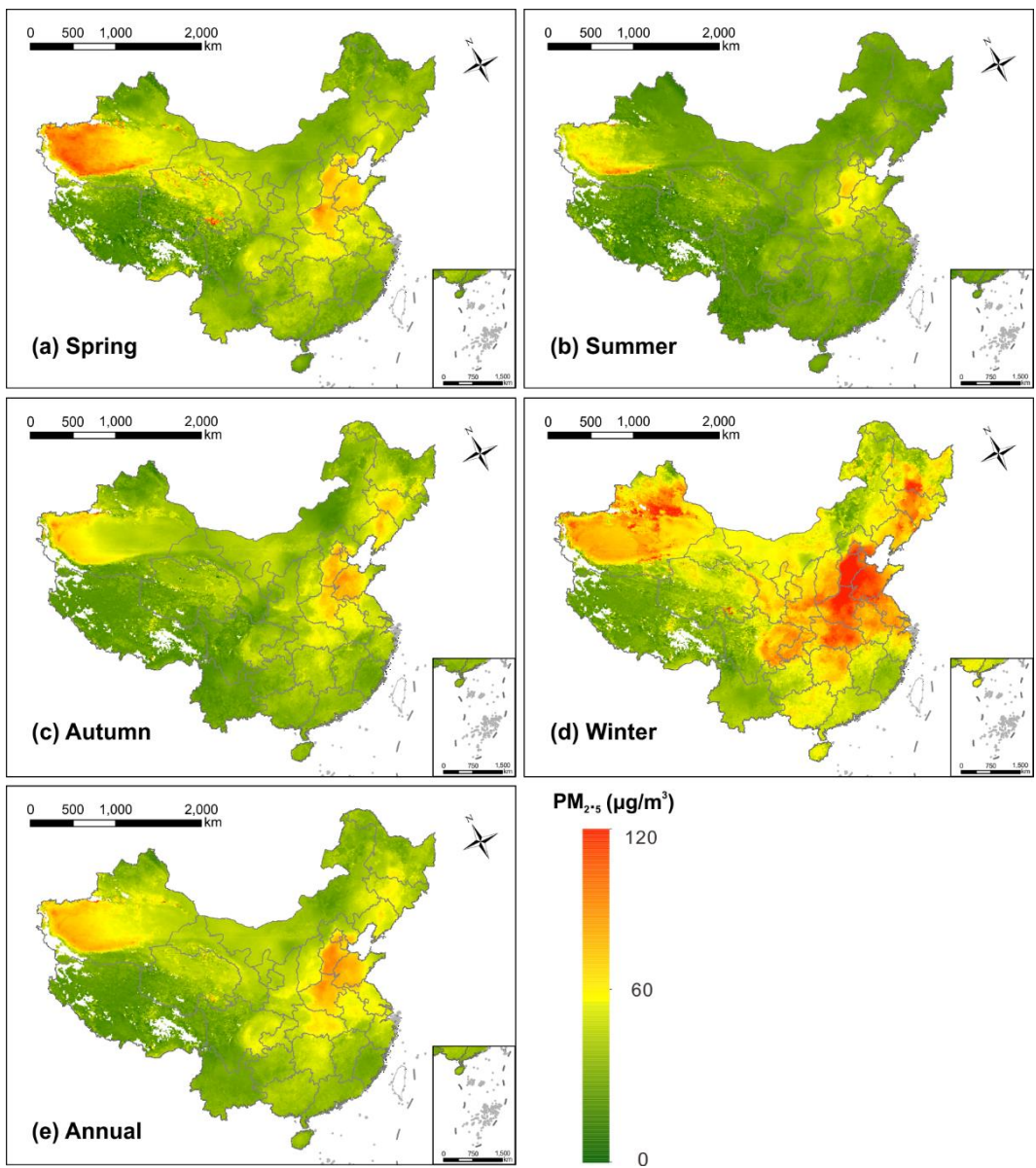

**Fig. 5**. Satellite-derived mapping of ground-level $PM_{2.5}$ concentrations over China in 2015. The white regions suggest missing data.

*4.5. Discussion*

To simultaneously consider nonlinearity and spatiotemporal heterogeneities of AOD-$PM_{2.5}$ relationship, GTWNNs were proposed to estimate ground-level $PM_{2.5}$ in this paper. Compared with traditional neural networks, the time-location-specific GTWNNs use a local fitting strategy, and need to address two essential issues. One is the amount of samples, the other is computation cost. For the first one, it should be noted that all samples on the estimation day and previous days are adopted for establishing a GTWNN model. Even though the samples with very small weights are excluded, the GRNN model, which was specifically designed for regression problems, is inherently effective to work with relatively limited samples. For the second issue, the GTWNNs are set up on every location and day, the computation cost seems to be notable enlarged compared with global neural networks. This challenge was addressed by two aspects of work. On the one hand, the samples with too small weights were exclude in the establishment of GTWNNs. On the other hand, the GRNN model used here has a fast solving ability so that amounts of iterations were avoided. Therefore, we hold the beliefs that the proposed GTWNNs model is an effective way for the assessment of ground-level $PM_{2.5}$ concentrations from satellite observations.

Up to now, many validation strategies have been adopted for the performance evaluation of AOD-$PM_{2.5}$ models, for example, sample-based cross-validation, site-based cross-validation, DOY (day of year)-based cross-validation (Ma et al., 2016b) and historical validation (Ma et al., 2016a), etc. These validation strategies often make the readers confused, because the same model may report different results with different validation strategies. Actually, the DOY-based cross-validation and historical validation are more suitable for the performance evaluation of temporal prediction. For satellite-based spatial mapping of PM$_{2.5}$, the most widely used are sample-based and site-based cross-validation. Compared with sample-based cross-validation, the site-based cross-validation is more reasonable to reflect the real prediction ability, because sample-based cross-validation may use the samples from the station itself on the previous time. While this is inconsistent with the actual situation for spatial prediction of PM$_{2.5}$. In this study, the GTWR model reports a relatively large decrease from sample-based cross-validation to site-based cross-validation, implying a relatively poorer ability for spatial prediction. However, the proposed GTWGRNN model achieves a site-based cross-validation R$^2$ of 0.79, indicating that this model can count for 79% of the variations of PM$_{2.5}$ in the spatial mapping.

**5. Conclusion and future work**

In this paper, geographically and temporally weighted neural networks were proposed to estimate ground-level PM$_{2.5}$ concentrations in China via the integration of satellite-derived AOD and station-measured PM$_{2.5}$. Through the simultaneous consideration of nonlinearity and spatiotemporal heterogeneities of AOD-PM$_{2.5}$ relationship, GTWNNs have achieved a satisfied performance for ground-level PM$_{2.5}$ estimation. The sample-based and site-based cross-validation R$^2$ values are 0.80 and 0.79, respectively. Comparisons with previous spatiotemporal models, i.e., D-GWR and GTWR, show that the GTWNNs model yields a great advantage. These results suggest that the proposed GTWNNs model is an effective way for estimating spatial PM$_{2.5}$ data with high resolutions. On this basis, the annual and seasonal distributions of PM$_{2.5}$ were accurately mapped in the whole of China. It is predicted that about 70% of grid cells have high PM$_{2.5}$ values of greater than CNNAQS level 2 standard (35 $\mu g/m^3$).

In future works, firstly, based on the proposed GTWNNs model, we are going to estimate spatial PM$_{2.5}$ concentrations using satellite AOD products with higher resolutions. For instance, the MODIS multi-Angle Implementation of Atmospheric Correction (MAIAC) AOD products with a resolution of 1 km (Lyapustin et al., 2011a; Lyapustin et al., 2011b), can provide more details for PM$_{2.5}$ estimates. In addition, the missing AOD data results in some challenges for the spatiotemporal models to fully capture the spatial and temporal variations of AOD-PM$_{2.5}$ relationship. Whether it is possible to develop an improved geographically and temporally weighted neural network model considering the missing data deserves future works.

**Acknowledgments**

This work was funded by the National Key R&D Program of China (No. 2016YFC0200900) and the National Natural Science Foundation of China (No. 41422108). We are grateful to the China National Environmental Monitoring Center (CNEMC, http://106.37.208.233:20035/), the Level 1 and Atmosphere Archive and Distribution System (LAADS, https://ladsweb.modaps.eosdis.nasa.gov/), and the Goddard Space Flight Center Distributed Active Archive Center (GSFC DAAC, http://gmao.gsfc.nasa.gov/GMAO_products/) for providing ground PM$_{2.5}$ measurements, satellite data, and meteorological data, respectively. The details of data used are listed in Section 2.